\documentclass{elsart}

\usepackage{natbib}

 \usepackage{epsf}

\usepackage{amssymb}

\begin{document}

\begin{frontmatter}



\title{On the coupling between different species during
recombination}


\author{Steen Hannestad}

\address{NORDITA, Blegdamsvej
17, DK-2100 Copenhagen, Denmark}

\begin{abstract}
Measurements of fluctuations in the Cosmic Microwave Background
Radiation (CMBR) is one of the most promising methods for measuring
the fundamental cosmological parameters. However, in order to
infer parameters from precision measurements it is necessary to
calculate the theoretical fluctuation spectrum to at least the
measurement accuracy. Standard treatments assume that electrons,
ions and neutral hydrogen are very tightly coupled during the
entire recombination history, and that the baryon-photon plasma
can be treated as a two-fluid system consisting of baryons and
photons interacting via Thomson scattering. We investigate the
validity of this approximation by explicitly writing down and
solving the full set of Boltzmann equations for electrons, ions,
neutral hydrogen and
photons. The main correction to the standard treatment is from
including Rayleigh scattering between photons and neutral hydrogen,
a change of less than 0.1\% in the CMBR power spectrum. 
Our conclusion is thus that the standard
treatment of the baryon-photon system is a very good approximation,
better than any possible measurement accuracy.

\end{abstract}


\end{frontmatter}


\section{introduction}

Anisotropies in the Cosmic Microwave Background Radiation (CMBR)
were first
detected in 1992 by the COBE satellite (Smoot et al.\ 1992). 
The amplitude and distribution
of these temperature fluctuations are closely related to the
underlying cosmological model. Thus, a precision measurement of the
CMBR fluctuations can in principle yield very precise information about
the values of the fundamental cosmological parameters, $\Omega,
\Omega_\Lambda, \Omega_b, H_0$ etc. 
(see for instance Bond et al. 1994, Jungman et al.\
1996a,b, Bond, Efstathiou \& Tegmark 1997, Eisenstein, Hu \& Tegmark 1999).  
Normally the fluctuations are
parametrized in terms of spherical harmonics as
\begin{equation}
\frac{\Delta T}{T}(\theta,\phi) =\sum_{lm}a_{lm}Y_{lm}(\theta,\phi),
\end{equation}
where the $a_{lm}$ coefficients are related to the power spectrum by
$C_l = \langle a^*_{lm} a_{lm} \rangle_m$.  For purely Gaussian
fluctuations the power spectrum contains all statistical information
about the fluctuations.
The data from COBE are not sufficiently accurate for a useful
determination of the cosmological parameters. However, a new
generation of high-precision experiments will be able to do this. Two
balloon experiments, BOOMERANG (de Bernardis et al. 2000)
and MAXIMA (Hanany et al. 2000), have already measured the
first and second acoustic peaks in the power spectrum (up to $l
\lesssim 800$).  In the next few years there will be data of even
better quality available from the MAP and PLANCK satellite experiments
\footnote{For information on these missions see the internet
pages for MAP (http://map.gsfc.nasa.gov) and PLANCK
(http://astro.estec.esa.nl/Planck/).}.
However, extracting precise information about the cosmological
parameters means that we should be able to reliably calculate the
theoretical fluctuation spectrum to at least as high precision as that
of the best measurement.

Calculating the CMBR power spectrum is a very complicated issue.  It
involves solving the coupled Boltzmann equations for all different
particle species present. This framework has been described in great
detail in many papers (see e.g.\ Ma \& Bertschinger 1995), 
and Hu, Scott, Sugiyama \& White (1995) for instance have discussed
the influence of various physical assumptions on the final CMBR
spectrum.

Since photons exchange momentum and
energy with the baryons, a core issue of any CMBR calculation is the
treatment of the photon-baryon system.
The photons mainly exchange energy with the baryons via Thomson
scattering on free electrons. The number of free electrons can be
calculated by solving the recombination equations. These equations
were first formulated by Peebles (1968)
and independently by Zeldovich, Kurt \& Sunyaev (1969).
Until recently this treatment was used in all calculations of CMBR
anisotropies, but it is only accurate to a few percent.  Recently
Seager, Sasselov \& Scott (1999a,b)
have provided a much more accurate method for
calculating recombination.

However, one assumption which has been made consistently in all
treatments is that electrons, protons and neutral hydrogen atoms are
very tightly coupled and can exchange energy and momentum much
faster than any other relevant timescale
(see e.g.\ Ma \& Bertschinger 1995).  This means that
the photon-baryon system can be treated in a two-fluid approximation,
where photons exchange energy with a baryon-fluid which is assumed to
have infinitely strong self-interaction.

In the present paper we calculate interaction and energy-momentum
exchange rates for the entire baryon-photon system, and write down the
full set of Boltzmann equations for the system of electron, ions,
neutrals and photons.  
Solving this system of equations we find that
the assumption of a
tightly coupled baryon fluid interacting with photons only through
Thomson scattering is a very good approximation, valid at the
$10^{-3}$ level.

In section 2 we calculate the relevant rates for different processes,
in section 3 we derive the Boltzmann equations for a multi-fluid
baryon plasma, and in section 4 we describe numerical results of
solving this extended set of equations.  Finally, section 5 contains a
discussion of our results.


\section{Reaction rates in the photon-baryon plasma}

\subsection{Photons}

{\it Electrons ---} The main energy exchange mechanism between photons
and electrons is Thomson scattering. The scattering rate per photon is
given by
\begin{equation}
\Gamma = n_e \langle \sigma_T v \rangle = 1.7 \times 10^{-8} \,\,
\Omega_b h^2 x_e T_{\rm eV}^3 \,\, {\rm s}^{-1},
\end{equation}
where $\sigma_T = 8 \pi \alpha^2/3 m_e^2$ is the Thomson scattering
cross-section.  Relativistic corrections to the standard cross section
are important at the $O(T/m_e) \simeq 10^{-5}$ level 
(Hu, Scott, Sugiyama \& White 1995).

The other main photon-electron processes are bremsstrahlung and double
Compton scattering.  The rates for these two processes are given
roughly by (Hu \& Silk 1993a,b, Lightman 1981)
\begin{equation}
\Gamma_{\rm BS} = 3 \times 10^{-15} \,\, T_{\rm eV}^{5/2} x_e^2
(\Omega_b h^2)^2 \,\, {\rm s}^{-1}.
\end{equation}
and
\begin{equation}
\Gamma_{\rm DC} = 2 \times 10^{-22} \,\, T_{\rm eV}^{5} x_e (\Omega_b
h^2) \,\, {\rm s}^{-1}.
\end{equation}

{\it Ions ---} Photons scatter on ions in exactly the same way as they
do on electrons.  However, because of the mass difference the cross
section is much smaller
\begin{equation}
\sigma = \sigma_T \left(\frac{m_e}{m_p}\right)^2 \simeq 3 \times
10^{-7} \,\, \sigma_T.
\end{equation}

{\it Neutrals ---} Photon scattering on neutrals (Rayleigh scattering)
is characterised by the cross section
(Jackson 1962, Lightman 1979, Peebles \& Yu 1970)
\begin{equation}
\sigma_R = \sigma_T (\omega_\gamma/\omega_0)^4, \,\,\, \omega_0=13.6
\,\,\, {\rm eV},
\label{eq:rayleigh}
\end{equation}
if $\omega \ll \omega_0$. 
This cross-section is energy dependent, but the thermal average
of the cross section leads to a scattering rate of
\begin{equation}
\Gamma_R = 2 \times 10^{-10} \,\, \Omega_b h^2
(1-x_e) T_{\rm eV}^7 \,\, {\rm s}^{-1},
\end{equation}
for $T_{\rm eV} \lesssim 1$ eV.   
However, close to the epoch of recombination, the photons are not
in scattering equilibrium (Hu \& Silk 1993a,b) and the different
photon modes do not exchange energy. Therefore the energy dependence 
of Rayleigh scattering can have different effects in different
wavelength bands. We will return to this question in section 5.

\subsection{Electrons}

{\it Photons ---} Thomson scattering gives a rate per electron of
\begin{equation}
\Gamma = n_\gamma \langle \sigma_T v \rangle \simeq 1.70 \,\, T_{\rm
eV}^3 \,\, {\rm s}^{-1}.
\label{eq:8}
\end{equation}

{\it Ions ---} Coulomb scattering on ions is characterized by the
cross section
\begin{equation}
\sigma_C = \frac{3}{2 v^4} \sigma_T \ln \Lambda,
\end{equation}
where $\Lambda$ is the Coulomb logarithm
\begin{equation}
\Lambda = \frac{3}{2}\left(\frac{T^3}{\pi \alpha^3 n_e (1+x_e)}\right)^{1/2}.
\end{equation}
The scattering rate is
then given by
\begin{equation}
\Gamma = n_e \langle \sigma_C v \rangle \simeq \frac{3}{2} \sigma_T
\ln \Lambda \left( \frac{m_e}{3 T} \right)^{3/2},
\end{equation}
and inserting numbers gives a rate of
\begin{equation}
\Gamma = 1.8 \, \log \Lambda \, \Omega_b h^2 x_e
T_{\rm eV}^{3/2} \,\, {\rm s}^{-1}.
\end{equation}

In addition to Coulomb scattering electrons and ions can recombine to
neutral hydrogen. The rate for this process is given roughly by
(Lightman 1979, Ma \& Bertschinger 1995)
\begin{equation}
\Gamma = 1.7 \times 10^{-7} \,\, \Omega_b h^2 x_e
T_{\rm eV}^{5/2} \,\, {\rm s}^{-1},
\end{equation}
which is entirely negligible as a means of momentum transfer.

{\it Neutrals ---} Electron scattering on hydrogen atoms at low energy
is primarily s-wave, so the cross-section is given by (Mott \& Massey 1965)
\begin{equation}
\sigma_N = \frac{4\pi}{k^2}\sum_{l=0}^{\infty} (2l+1) \sin^2 \delta_l
\simeq \frac{4\pi}{k^2} \sin^2 \delta_{l=0}
\end{equation}

The s-wave phase shift has been calculated for instance by Schwartz
(1961).
To a reasonable approximation we may put
$\delta \simeq \pi$ for $E \ll \omega_0$.  In that case the rate is
\begin{equation}
\Gamma = n_n \langle \sigma_N v \rangle \simeq n_n \left\langle
\frac{4\pi}{2 m_e E_e} \left(\frac{2
E_e}{m_e}\right)^{1/2}\right\rangle
\end{equation}
Inserting numbers into the above equation yields the rate
\begin{equation}
\Gamma = 0.16 \,\, \Omega_b h^2 (1-x_e) T_{\rm
eV}^{5/2} \,\, {\rm s}^{-1}.
\label{eq:nscatter}
\end{equation}

\subsection{Ions}

We have already seen that Thomson scattering is negligible for
protons.  The Coulomb scattering rate is the same as for electrons.
Scattering on neutrals is very inefficient because of the much higher
mass of protons compared with electrons.  Finally, the recombination
rate is the same for protons as it is for electrons.

\subsection{Neutral hydrogen}

We shall in the present paper assume that all neutral hydrogen is in
the ground state, an approximation which is very good for our
purposes.

{\it Photons ---} The Rayleigh scattering rate has been calculated
above to be
\begin{equation}
\Gamma_R = n_\gamma \langle \sigma_R v \rangle
\simeq 6.4 \times 10^{-3} T_{\rm eV}^7 \,\,\, {\rm s}^{-1}
\label{eq:17}
\end{equation}

Photo-ionization of neutral hydrogen could in principle also be
important. The rate for this process is given by (Ma \& Bertschinger 1995)
\begin{equation}
\Gamma = \frac{32}{\sqrt{54} \pi} m_e^{-1/2} T^{3/2} e^4
\left(\frac{T}{B}\right) \phi(T),
\end{equation}
where $B$ is the binding energy of the given level and $\phi(T) =
0.448 \ln(B/T)$. For ionization from the ground state this is
approximately
\begin{equation}
\Gamma = 9.1 \times 10^{10} \,\, T_{\rm eV}
e^{-13.6/T_{\rm eV}} \,\, {\rm s}^{-1}.
\end{equation}
Finally, neutral hydrogen can also absorb momentum from the photon gas
via photo-excitation. During recombination the Ly-$\alpha$ line is
densely populated by resonance photons (Peebles 1968). 
This effective non-thermal distribution function across the line
we denote with $f_\alpha$. In terms of this the photo-excitation rate
from the ground state is
\begin{equation}
\Gamma = 2.1 \times 10^5 f_\alpha T_{\rm eV}^{-1/2} \,\, {\rm s}^{-1}.
\end{equation}
From the recombination calculation we find $f_\alpha$ and thus
$\Gamma$.  The system of equations we use is that provided by
Peebles (Peebles 1968). 
Although it is not as accurate as that of Seager, Sasselov
\& Scott (1999a,b), 
it is adequate for our purpose.  Note that photo-excitation
does not in itself transfer any momentum between hydrogen and the
other species. Almost all photons in the line are resonance photons
and not thermal background photons. However, hydrogen in excited states
is ionized almost completely and so the above rate can be seen
effectively as the rate for converting hydrogen and $ep$.

{\it Electrons ---} Scattering on electrons is by far the most
important means of exchanging momentum with other species. The rate is
roughly given by (c.f.\ Eq.~(\ref{eq:nscatter}))
\begin{equation}
\Gamma  = n_e \langle \sigma_N v \rangle \simeq 
0.16 \,\, \Omega_b h^2 x_e T_{\rm eV}^{5/2}
\,\, {\rm s}^{-1}.
\end{equation}

{\it Ions ---} Scattering on ions is by s-wave scattering exactly as
for electrons. However, the cross
section is much smaller because of the much higher mass of protons.

\subsection{Ion-electron fluid}

The Coulomb interaction between electrons and ions is already the
dominant reaction rate in the plasma. However, as soon as there is any
motion of electrons relative to ions, an electric field quickly builds
up and acts in the direction opposite to the motion
(Hogan 2000). Thus, electrons and ions are even more tightly coupled
than the Coulomb scattering suggests and effectively this means that
electrons and protons can be treated as one tightly coupled species.

\subsection{Comparison of rates}

From the above equations we can compare the different rates and
identify the dominant ones for momentum exchange between the different
species. However, the above equations describe the scattering rates,
not the rates for momentum transfer between different species.
These two quantities are to a good approximation equivalent, except 
in the case where photons scatter on massive particles.
Because photons have very low momenta compared with massive particles,
they are inefficient at transferring momentum to these particles. 
In each scattering, the photon on average loses a large fraction of its
momentum, whereas the massive particle only loses a fraction $\sim T/m$
of its momentum. This effect is only important for the processes 
described in Eqs.~(\ref{eq:8}) and (\ref{eq:17}). 
Both these equations should be
multiplied by a factor $T/m_p$ to find the rate of momentum transfer.
When this is done, all rates are momentum transfer rates and can be 
directly compared.

Fig.~1 shows the different rates as a function of time.
From the figure it is clear which processes we need to include, they
are: \\ 1) Thomson scattering which couples photons with the
ion-electron plasma \\ 2) Rayleigh scattering which couples photons
with neutral hydrogen \\ 3) s-wave scattering which couples the
ion-electron fluid with neutral hydrogen (this process is dominant for
$z \lesssim 2000$).

\begin{figure}
\begin{center}
\epsfysize=12truecm\epsfbox{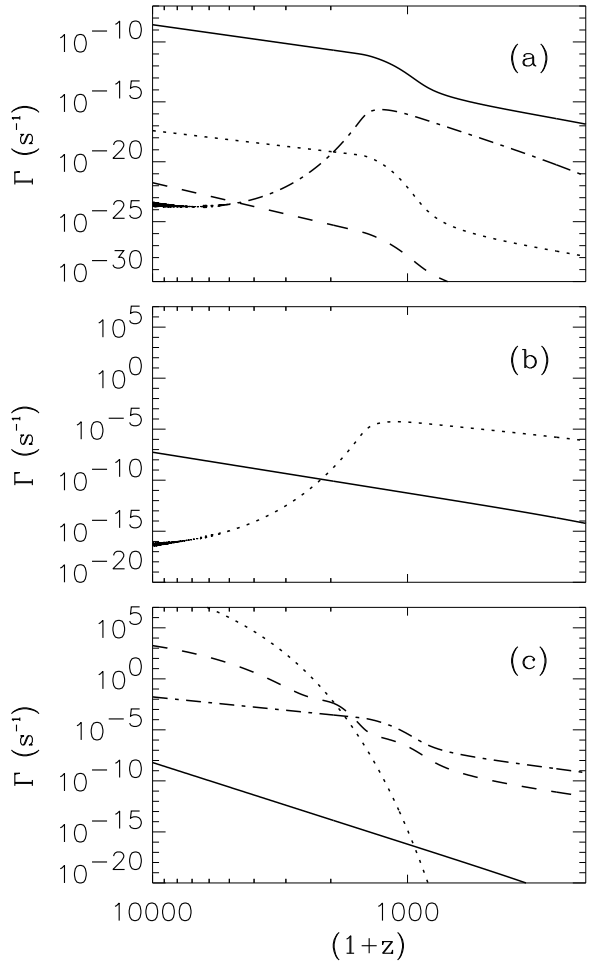}
\vspace*{0.5truecm}
\end{center}
\baselineskip 17pt
\caption{Momentum transfer rates for the different species present 
during CMBR formation: (a) photons, (b) ion-electron fluid,
(c) neutrals. In each case, the labeling is as follows:
(a) Full line is Thomson scattering, dotted is bremsstrahlung,
dashed is double Compton scattering and dot-dashed is Rayleigh scattering.
(b) Full line is Thomson scattering, dotted is scattering on neutrals.
(c) Full line is Rayleigh scattering, dotted is photo-ionization,
dashed is photo-excitation and dot-dashed is scattering on electrons.
All rates were calculated assuming a standard flat CDM model with
$\Omega_m=1, \Omega_b=0.05, H_0=50 \,\, {\rm km \, s^{-1} \,
Mpc^{-1}}$.}
\label{fig1}
\end{figure}


\section{Boltzmann equations for a multi-fluid baryon-photon plasma}

The evolution of different particle species can be described via the
Boltzmann equation. In deriving the equations below we shall work in
synchronous gauge because the numerical code for calculating CMBR
power spectra, CMBFAST (Seljak \& Zaldarriaga 1996), 
is written in this gauge.  As the time
variable we use conformal time $d \tau \equiv dt/a(t)$, where $a(t)$
is the scale factor. Finally, instead of using physical momentum,
$p_j$, we work with comoving momentum, $q_j = a p_j$, because it is a
conserved quantity in the expanding universe. Finally, we parametrize
it as $q_j = q n_j$, where $q$ is the magnitude and $n_j$ is a
3-vector describing its direction.  Generically, the Boltzmann
equation can always be written as
\begin{equation}
L[f] = \frac{Df}{D\tau} = C[f],
\end{equation}
where $L[f]$ is the Liouville operator. 
The collision operator on the right-hand side describes
any possible collisional interactions.

We then write the distribution function as
\begin{equation}
f(x^i,q,n_j,\tau) = f_0(q) [1+\Psi(x^i,q,n_j,\tau)],
\end{equation}
where $f_0(q)$ is the unperturbed distribution function.
$f_0$ can be found by solving the unperturbed Boltzmann equation
(Kaplinghat et al. 1999)
\begin{equation}
\frac{df_0}{d\tau} = C[f_0].
\end{equation}
The perturbed part of the Boltzmann equation can be written as
an evolution equation for $\Psi$ in $k$-space (Ma \& Bertschinger 1995)
\begin{equation}
\frac{1}{f_0} L[f] = \frac{\partial \Psi}{\partial \tau} + i \frac{q}{\epsilon}
\mu \Psi + \frac{d \ln f_0}{d \ln q} \left[\dot{\eta}-\frac{\dot{h}+6\dot{\eta}}
{2} \mu^2 \right] = \frac{1}{f_0} C[f],
\end{equation}
where $\mu \equiv n^j \hat{k}_j$.
$h$ and $\eta$ are the metric perturbations, defined from the perturbed space-time
metric in synchronous gauge (Ma \& Bertschinger 1995)
\begin{equation}
ds^2 = a^2(\tau) [-d\tau^2 + (\delta_{ij} + h_{ij})dx^i dx^j],
\end{equation}
\begin{equation}
h_{ij} = \int d^3 k e^{i \vec{k}\cdot\vec{x}}\left(\hat{k}_i \hat{k}_j h(\vec{k},\tau)
+(\hat{k}_i \hat{k}_j - \frac{1}{3} \delta_{ij}) 6 \eta (\vec{k},\tau) \right).
\end{equation}

{\it Collisionless Boltzmann equation ---}
At first we assume that $\frac{1}{f_0} C[f] = 0$.
We then expand the perturbation as 
\begin{equation}
\Psi = \sum_{l=0}^{\infty}(-i)^l(2l+1)\Psi_l P_l(\mu).
\end{equation}
One can then write the collisionless
Boltzmann equation as a moment hierarchy for the $\Psi_l$
by performing the angular integration of $L[f]$
\begin{eqnarray}
\dot\Psi_0 & = & -k \frac{q}{\epsilon} \Psi_1 + \frac{1}{6} \dot{h} \frac{d \ln f_0}
{d \ln q} \label{eq:psi0}\\
\dot\Psi_1 & = & k \frac{q}{3 \epsilon}(\Psi_0 - 2 \Psi_2) \label{eq:psi1}\\
\dot\Psi_2 & = & k \frac{q}{5 \epsilon}(2 \Psi_1 - 3 \Psi_3) - \left(\frac{1}{15}
\dot{h}+\frac{2}{5}\dot\eta\right)\frac{d \ln f_0}{d \ln q} \\
\dot\Psi_l & = & k \frac{q}{(2l+1)\epsilon}(l \Psi_{l-1} - (l+1)\Psi_{l+1}) 
\,\,\, , \,\,\, l \geq 3
\end{eqnarray}
It should be noted here that the first two hierarchy equations are directly
related to the energy-momentum conservation equation.
This can be seen in the following way. Let us define the density and
pressure perturbations of the dark matter fluid as (Ma \& Bertschinger 1995)
\begin{eqnarray}
\delta & \equiv & \delta \rho/\rho \\
\theta & \equiv & i k_j \delta T^0_j/(\rho+P) \\
\sigma & \equiv & -(\hat{k}_i \hat{k}_j - \frac{1}{3} \delta_{ij})
(T^{ij}-\delta^{ij}T^k_k/3).
\end{eqnarray}
Then energy and momentum conservation implies that (Ma \& Bertschinger 1995)
\begin{eqnarray}
\dot\delta & = & -(1+\omega)\left(\theta+\frac{\dot h}{2}\right)-
3 \frac{\dot a}{a} \left(\frac{\delta P}{\delta \rho} - \omega \right) \delta 
\label{eq:energy}\\
\dot \theta & = & -\frac{\dot a}{a} (1-3 \omega)\theta - \frac{\dot \omega}{1+\omega}
\theta + \frac{\delta P/ \delta \rho}{1+\omega} k^2 \delta - k^2 \sigma.
\label{eq:mom}
\end{eqnarray}
By integrating Eq.~(\ref{eq:psi0}) over $q^2 \epsilon dq$,
one gets Eq.~(\ref{eq:energy}) and by integrating Eq.~(\ref{eq:psi1})
equation over $q^3 dq$ one retrieves Eq.~(\ref{eq:mom}).

{\it Collisional Boltzmann equation ---}
The baryon-photon system is coupled by interactions, 
so $C[f] \neq 0$.
In the standard treatment, where all baryons are assumed to be infinitely
tightly coupled to each other, there are only two Boltzmann equations,
one for the photons and one for the baryons (Ma \& Bertschinger 1995).
The baryons are highly non-relativistic and we need only consider the
first two terms in the Boltzmann hierarchy, corresponding to energy
and momentum. 
For photons many more terms need to be incorporated. However, in the present
treatment we suppress writing out explicitly these higher order terms.
The collision term
for the system has been calculated many times in the literature
and the collisional Boltzmann equations are (Ma \& Bertschinger 1995)

{\it Baryons:}
\begin{eqnarray}
\dot\delta & = & -\left(\theta+\frac{\dot h}{2}\right) \\
\dot \theta & = & -\frac{\dot a}{a} \theta - 
\frac{\delta P}{\delta \rho} k^2 \delta 
+ \frac{4 \rho_\gamma}{3 \rho_b} a n_e \sigma_T (\theta_\gamma - \theta_b).
\end{eqnarray}

{\it Photons:}
\begin{eqnarray}
\dot\delta & = & -\frac{4}{3}\left(\theta+\frac{\dot h}{2}\right) \\
\dot \theta & = & k^2 (\delta_\gamma/4-\sigma_\gamma)
+ a n_e \sigma_T (\theta_b - \theta_\gamma) \\
& & \hspace{-0.5cm} + \,\,\, {\rm higher \,\,\, order \,\,\, terms} 
\end{eqnarray}
The full hierarchy can be found for instance in Ma and Bertschinger
(1995).

In our case, instead of an infinitely tightly coupled baryon-electron
fluid we have three interacting ``species'': electrons ($e$), 
ions ($i$) and neutrals ($n$). This means that 4 different Boltzmann
hierarchies have to be solved simultaneously. As discussed in section 3,
the electrons and ions can be considered as infinitely tightly coupled
because of charge neutrality. This ion-electron fluid we denote
by the subscript $i-e$. The Boltzmann hierarchy is now

{\it Ion-electron fluid:}
\begin{eqnarray}
\dot\delta & = & -\left(\theta+\frac{\dot h}{2}\right) 
+ \frac{\dot\rho_{i-e}}{\rho_{i-e}} (\delta_n - \delta_{i-e}) \\
\dot \theta & = & -\frac{\dot a}{a} \theta - 
\frac{\delta P}{\delta \rho} k^2 \delta 
+ \frac{4 \rho_\gamma}{3 \rho_{i-e}} a n_e \sigma_T 
(\theta_\gamma - \theta_{i-e}) + a n_n 
\langle \sigma_N v \rangle (\theta_n - \theta_{i-e}).
\end{eqnarray}

{\it Neutrals:}
\begin{eqnarray}
\dot\delta & = & -\left(\theta+\frac{\dot h}{2}\right) 
+ \frac{\dot\rho_{n}}{\rho_{n}} (\delta_{i-e} - \delta_{n}) \\
\dot \theta & = & -\frac{\dot a}{a} \theta + 
\frac{\delta P}{\delta \rho} k^2 \delta 
+ \frac{4 \rho_\gamma}{3 \rho_{n}} a n_n \langle \sigma_R \rangle
(\theta_\gamma - \theta_{n}) + a n_{i-e} 
\langle \sigma_N v \rangle (\theta_{i-e} - \theta_n).
\end{eqnarray}

{\it Photons:}
\begin{eqnarray}
\dot\delta & = & -\frac{4}{3}\left(\theta+\frac{\dot h}{2}\right) \\
\dot \theta & = & k^2 (\delta_\gamma/4-\sigma_\gamma)
+ a n_e \sigma_T (\theta_{i-e} - \theta_\gamma)
+ a n_s \langle \sigma_R \rangle (\theta_n - \theta_\gamma)\\
& & \hspace{-0.5cm} + \,\,\, {\rm higher \,\,\, order \,\,\, terms} 
\end{eqnarray}

Here $\langle \sigma_R \rangle$ denotes the thermally averaged 
Rayleigh cross section.
In the equations for $\dot \delta$ for neutrals and ions, there is a new
term appearing because ions and neutrals can interconvert.
$\dot\rho$ can be found from solving the unperturbed Boltzmann equation
(Kaplinghat et al. 1999),
which in the present case is the recombination equation.


\section{Numerical results}

Solving the above system of equations is quite complicated, but we can 
easily estimate the importance of the different terms.
The first important term is the interaction between neutrals and
electrons/ions. We can estimate the relative velocity difference between
these two components as
\begin{equation}
\frac{\theta_n - \theta_{i-e}}{\theta_n}
\simeq \frac
{\frac{4 \rho_\gamma}{3 \rho_{i-e}} n_e \sigma_T}
{n_e \langle \sigma_N v \rangle}
\simeq 1 \times 10^{-7} T_{\rm eV}^{-3/2}
\end{equation}
Thus, the relative velocity building up between neutrals and 
electrons/ions is at most of $O(10^{-7})$. 
This correction is exceedingly
small and we can to a very good approximation
treat neutrals and electrons-ions as infinitely
tightly coupled. In that case the above system of equations almost 
reduces to that in the standard treatment

{\it Baryons:}
\begin{eqnarray}
\dot\delta & = & -\left(\theta+\frac{\dot h}{2}\right) \\
\dot \theta & = & -\frac{\dot a}{a} \theta - 
\frac{\delta P}{\delta \rho} k^2 \delta 
+ \frac{4 \rho_\gamma}{3 (\rho_{i-e}+\rho_n)} 
a (n_e \sigma_T + n_n \langle \sigma_R \rangle) (\theta_\gamma - \theta).
\end{eqnarray}

{\it Photons}
\begin{eqnarray}
\dot\delta & = & -\frac{4}{3}\left(\theta+\frac{\dot h}{2}\right) \\
\dot \theta & = & k^2 (\delta_\gamma/4-\sigma_\gamma)
+ a (n_e \sigma_T + n_n \langle \sigma_R \rangle) 
(\theta_b - \theta_\gamma) \\
& & \hspace{-0.5cm} + \,\,\, {\rm higher \,\,\, order \,\,\, terms} 
\end{eqnarray}
Except for the fact that Rayleigh scattering is now included this form is
identical to the standard treatment. Note that the polarization terms
for photons are different for Rayleigh scattering compared with
for Thomson
scattering because Rayleigh scattering corresponds to scattering
in a dipole field (see e.g.\ Jackson 1962). 
However, this is a higher order effect which is
we shall not discuss further in the present paper.

Thus, the corrections to the standard treatment come almost solely
from including Rayleigh scattering in the calculations of the
temperature power spectrum.
We have modified the CMBFAST code to include the effect of Rayleigh
scattering. In Fig.\ 2 we show how the CMBR temperature power spectrum
changes due to this inclusion.
The difference between the two spectra increases with $l$, but is
always below $10^{-3}$. 
Since the absolute accuracy of CMBFAST is only about 1\%, one might
worry that the effect is due to numerical noise. However, the effect
shown in Fig.~2 is a differential effect, and the accuracy of CMBFAST
should therefore be substantially better.
\begin{figure}
\begin{center}
\epsfysize=12truecm\epsfbox{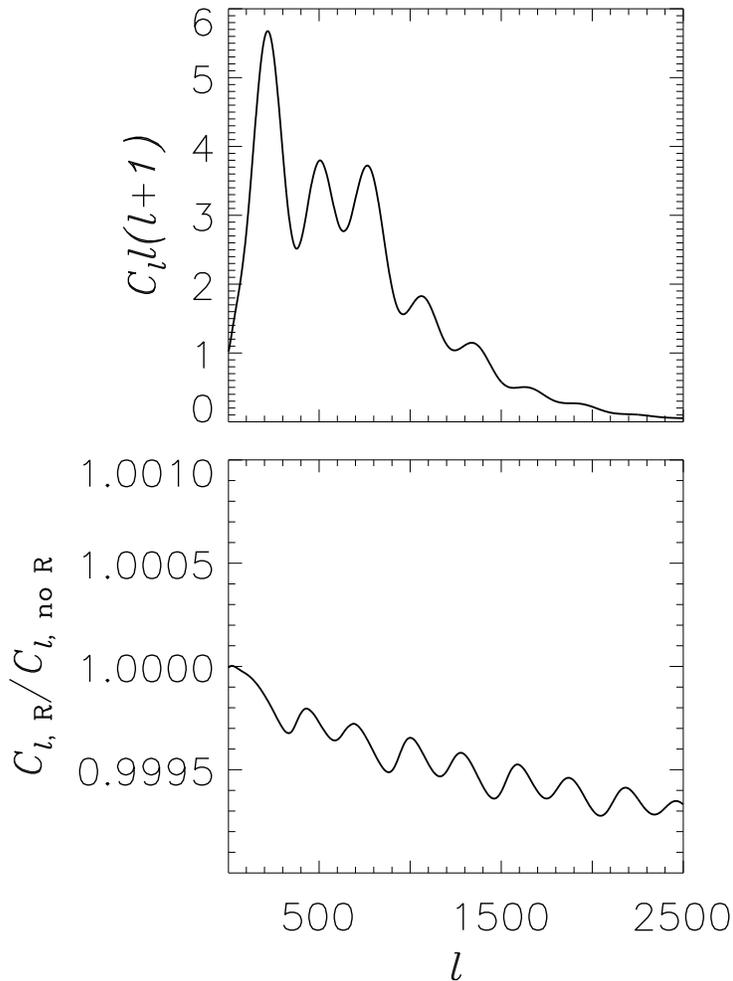}
\vspace*{0.5truecm}
\end{center}
\baselineskip 17pt
\caption{CMBR power spectra for a standard flat CDM model with 
$\Omega_m=1, \Omega_b=0.05, H_0=50 \,\, {\rm km \, s^{-1} \,
Mpc^{-1}}$, calculated with (full line) and without (dotted line)
the inclusion of Rayleigh scattering.
The upper panel shows the actual power spectra which are
so close that they are indistinguishable on the figure. The lower panel
shows the ratio of the two power spectra.}
\label{fig2}
\end{figure}

It is in fact possible from analytic arguments, to understand how
the inclusion of Rayleigh
scattering dampens the fluctuation spectrum.
Three effects contribute to change the CMBR power spectrum:
1) Diffusion damping of the perturbations close to recombination
is changed by including Rayleigh scattering,
2) The last scattering surface is moved to slightly lower redshift,
so that $\rho_m/\rho_\gamma$ is higher. This means that the early ISW
effect is less important, and suppresses power around the horizon
size at recombination.
3) Since the last scattering surface is moved to lower redshift,
$R \equiv 3\rho_b/4\rho_\gamma$, is also higher. This parameter changes
the amplitude of the acoustic oscillations (Hu 1995).

The first effect is the most important and can to a reasonable
precision be calculated analytically. The diffusion damping of the
$k$-space fluctuation spectrum can be written roughly as (Hu 1995)
\begin{equation}
P(k)_{\rm damping} \simeq D^2(k) P(k)_{\rm no \,\, damping},
\end{equation}
where
\begin{equation}
{D}(k) = \int_0^{\eta_0} d \eta \dot\tau e^{-\tau}
e^{-(k/k_D)^2}.
\end{equation}
Here, $\dot\tau = \Gamma$ is the differential optical depth,
and $k_D \simeq \int d \eta/\dot\tau$ is the diffusion scale for a
photon. $\eta$ is the conformal time coordinate, $d\eta \equiv dt/a$. 

We then use the approximate relation that 
$l \simeq 2 k/H_0$ to write that
\begin{equation}
\left(\frac{C_{l, {\rm R}}}{C_{l, {\rm no \,\, R}}}\right)_{l \simeq 2 k/H_0}
\simeq \left(\frac{\int_0^{\eta_0} d \eta \dot\tau_{R} e^{-\tau_R}
e^{-(k/k_{D,R})^2}}
{\int_0^{\eta_0} d \eta \dot\tau_{\rm no \,\, R} e^{-\tau_{\rm no \,\, R}}
e^{-(k/k_{D, {\rm no \,\, R}})^2}}\right)^2
\label{eq:damp}
\end{equation}
\begin{figure}
\begin{center}
\epsfysize=7truecm\epsfbox{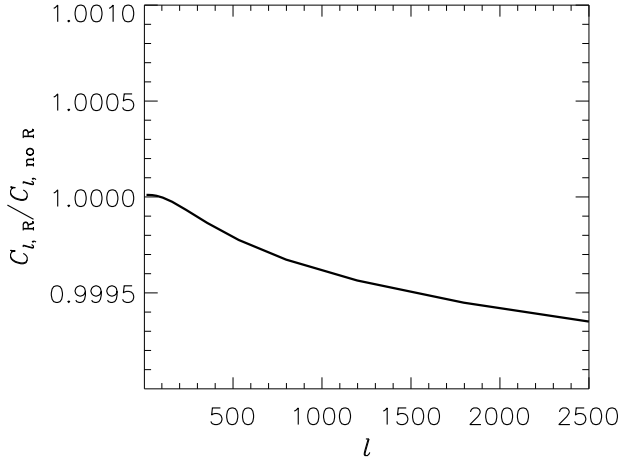}
\vspace*{0.5truecm}
\end{center}
\baselineskip 17pt
\caption{The damping of the CMBR power spectrum as a function of $l$,
calculated from the analytic estimate of Eq.~(\ref{eq:damp}).}
\label{fig3}
\end{figure}

This damping factor is shown as a function of $l$ in Fig.~3. It quite
nicely agrees with both the magnitude and shape of the damping seen
in the numerically calculated spectrum (Fig.~2).
The oscillatory behaviour of the damping in Fig.~2 comes from the
fact that the parameter $R$ is changed 
(Hu 1995, Landau, Harari and Zaldarriaga 2000), 
so that the oscillation amplitudes
are different when Rayleigh scattering is included.

Thus, the behaviour of the fluctuation damping due to Rayleigh scattering,
seen in Fig.~2,  
is understandable in terms of simple physical arguments.


\section{Discussion}

We have thoroughly reviewed the interactions between different particle
species in the baryon-photon plasma during recombination, as well as 
written down the complete set of Boltzmann equations for the system
of photons, electrons, ions and neutrals. 
In the standard CMBR calculations, the entire baryon plasma is treated
as infinitely tightly coupled to itself. We find that the error introduced
by this assumption is only at the $10^{-7}$ level. The main error in
the standard treatment is that Rayleigh scattering between photons
and neutral hydrogen is neglected. However, 
the difference in the power spectrum
when this process is included is at most of $O(10^{-3})$, even at high
$l$.

The benchmark for any CMBR calculation is the best possible precision
with which any measurement can be made. Because the ensemble average
needed in calculating $C_l$ in practise has to be replaced with an
average over $m$ there is an uncertainty in $C_l$ of
\begin{equation}
\frac{\sigma(C_l)}{C_l} \geq \sqrt{\frac{2}{2l+1}},
\end{equation}
which corresponds to the best precision any CMBR measurement can 
be made to.
Even at $l=2000$ this ``cosmic variance''
is $\sigma(C_l)/C_l  = 0.022$, which is an order 
of magnitude higher than the error introduced by neglecting Rayleigh
scattering, and several orders of magnitude larger than the error
introduced by treating neutrals and ions/electrons as infinitely
tightly coupled.
Notice, however, that this calculation has been performed using
a thermally averaged Rayleigh cross section. Since photons with 
different energy do not exchange energy at recombination
(Hu \& Silk 1993a,b), this different Rayleigh interaction rate
can lead to different fluctuation spectra in different wavelength
bands. A full calculation of this effect requires solving the
momentum dependent Boltzmann equation for the photon gas
(Ma \& Bertschinger 1995), and is beyond the scope of the present work.
However, one can get a rough feeling for how the effect of Rayleigh
scattering scales with photon energy.
\begin{figure}
\begin{center}
\epsfysize=7truecm\epsfbox{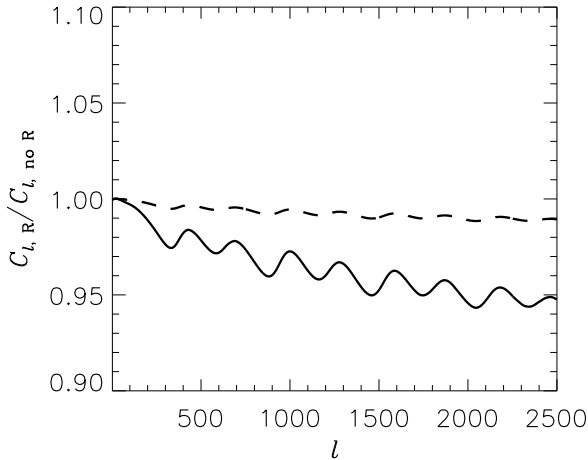}
\vspace*{0.5truecm}
\end{center}
\baselineskip 17pt
\caption{The damping of the CMBR power spectrum as a function of $l$,
calculated for the cases of $\sigma_R = 2^4 \langle \sigma_R \rangle$
(dashed line) and $\sigma_R = 3^4 \langle \sigma_R \rangle$
(solid line).}
\label{fig4}
\end{figure}
In Fig.~4 we show the effect on the power spectrum in the case where
the Rayleigh scattering is increased by factors of $2^4$ and $3^4$
above the value for the average photon energy. 
In this case, the effect
can be very significant and above the detection threshold.
To compare with observations,
the highest frequency channel of the Planck surveyor 
is at 857 GHz (see http://astro.estec.esa.nl/Planck/), 
which corresponds to $E_\gamma \simeq 15 T_0 \simeq 
5 \langle E_\gamma \rangle$
\footnote{Note that even though this frequency would from 
Eq.~(\ref{eq:rayleigh}) correspond to $\sigma \simeq 5^4 \langle \sigma
\rangle$, Eq.~(\ref{eq:rayleigh}) does not apply at such high energy
during recombination. The upper limit for the cross section would rather
be $\sigma \simeq \sigma_T$, which corresponds better to the case
of $\sigma = 3^4 \langle \sigma \rangle$, shown in Fig.~4.}.
So for the Planck surveyor it is not
impossible that the effect at high frequencies could be detectable.
However, this simplistic analysis clearly overestimates the effect,
because increasing the general Rayleigh scattering by a certain amount
implicitly assumes that all photons have more energy,
which is not really the case. A real calculation
using the momentum dependent Boltzmann equation would likely find a 
somewhat smaller effect, which could nevertheless still be significant
at high energy.
A second point is that the high frequency bands are subject to heavy
foreground contamination (Tegmark et al. 2000), 
so that a cosmological signal would probably not be visible.

\section*{Acknowledgements}

Use of the CMBFAST package for calculating CMBR anisotropies 
(Seljak \& Zaldarriaga 1996) is acknowledged.

\end{document}